\documentclass[showpacs]{elsarticle} 
\usepackage{amssymb,amsmath,amsthm,graphicx,amscd}
\usepackage[mathscr]{eucal}
\usepackage{enumerate,color,ulem}

\journal{Physics Letter B}
\begin{document}

\baselineskip=18pt
\allowdisplaybreaks

\begin{frontmatter}

\title{``Hot Entanglement"?\\-- A Nonequilibrium Quantum Field Theory Scrutiny}

\author[add]{J.-T. Hsiang}
\ead{cosmology@gmail.com}
\author[add,secadd]{B.~L.~Hu}
\ead{blhu@umd.edu}
\address[add]{Center for Particle Physics and Field Theory, Department of Physics, \\ Fudan University, Shanghai 200433, China}
\address[secadd]{Maryland Center for Fundamental Physics and Joint Quantum Institute, \\ University of Maryland, College Park, Maryland 20742, USA}

\begin{abstract}
The possibility of maintaining entanglement in a quantum system at finite, even high, temperatures -- the so-called `hot entanglement' -- has obvious practical interest, but also requires closer theoretical scrutiny.  Since quantum entanglement in a system evolves in time and is continuously subjected to environmental degradation, a nonequilibrium description by way of open quantum systems is called for. To identify the key issues and the contributing factors that may permit `hot entanglement' to exist, or the lack thereof, we carry out a model study of two spatially-separated, coupled oscillators in a shared bath depicted by a finite-temperature scalar field. From the Langevin equations we derived for the normal modes and the entanglement measure constructed from the covariance matrix we examine the interplay between direct coupling, field-induced interaction and finite separation on the structure of late-time entanglement. We show that the coupling between oscillators plays a crucial role in sustaining entanglement at intermediate temperatures and over finite separations. In contrast, the field-induced interaction between the oscillators which is a non-Markovian effect, becomes very ineffective at high temperature. We determine the critical temperature above which entanglement disappears to be bounded in the leading order by the inverse frequency of the center-of-mass mode of the reduced oscillator system, a result not unexpected, which rules out hot entanglement in such settings.
\end{abstract}

\begin{keyword}
\texttt{quantum field theory} \sep \texttt{nonequilibrium quantum dynamics} \sep \texttt{quantum open systems}\sep \texttt{thermal entanglement}

\PACS 03.65.Ud \sep 03.65.Yz \sep 03.67.-a
\end{keyword}

\end{frontmatter}

\section{Introduction}

Recently Galve et al~\cite{GalvePRL,EP} pointed out the possibility of keeping quantum entanglement alive in a system at high temperatures by driving the system of two oscillators with a time-dependent interaction term. This is important in practical terms because if entanglement in a quantum open system can be maintained at high temperatures, it eases the way how  devices for quantum information processing can be conceptualized and designed.  From a theoretical viewpoint understanding the basic mechanisms of obtaining this so-called `hot entanglement' \cite{Vedral} is also of great interest.

Before beginning the analysis, we note the word `hot'  conveys three layers of meaning in three different contexts, referring to quantum systems  A)  kept in thermal \textit{equilibrium} at all times,  B) in a \textit{nonequilibrium }condition and evolving, possibly but not necessarily, toward an equilibrium state,  and  C) in a \textit{nonequilibrium steady state} at late times.
In this study we derive the fully nonequilibrium dynamics of a system of two coupled quantum harmonic oscillators interacting with a common bath described by a  bosonic field at finite temperature $T$.   Thus our present work falls under Case B, which is in contrast to Case A \cite{AndWin,Anders}, where a quantum system is assumed to be already in equilibrium and remains that way. We depict how entanglement of the open quantum system evolves in time and  derive the critical temperature above which entanglement cannot survive. In an accompanying paper \cite{HotEnt2} we study one subcase of Case C,  that of a quantum system in nonequilibrium steady state (NESS) at late times, using the framework and results obtained in \cite{HHNESS}. The system we analyze there consists of two coupled quantum harmonic oscillators each interacting with its own bath, described by a scalar field, set at two different temperatures $T_1 > T_2$ which together form the environment. Carrying out a fully systematic analysis of how quantum entanglement in open systems under different nonequilibrium conditions evolves is, in our view, a necessity before any claim of ``hot entanglement" can be asserted.

In terms of methodology our present study makes use of the conceptual framework of quantum open systems \cite{qos} and the techniques of nonequilibrium quantum field theory \cite{CH08}.  It is a finite temperature generalization of  our recent work \cite{QEnt0} where the entanglement behavior at late times between two coupled and spatially separated oscillators interacting with a common bath modeled by a scalar field at zero temperature is analyzed in detail.  That work in turn is a generalization of the paper of Lin and Hu \cite{LH09} with coupling between the two oscillators added in the consideration.

\section{System Setup}

Our system is made up of two spatially separated coupled detectors,  which are entities with internal degrees of freedom (idf) $\chi_{1,2}$.  The idf of each detector is described by a harmonic oscillator of mass $m$ and bare frequency $\omega_{b}$.  This system is placed in a common finite-temperature bath modeled by a massless scalar field $\phi$ initially prepared in a thermal state at temperature $\beta^{-1}$. The system is allowed to interact with the bath initially at $t=0$. We want to track down its evolution in time, derive the entanglement dynamics between the two detectors  at late times and determine the critical temperature above which entanglement no longer exists.

The action of the whole system is
\begin{align}
	 S[\chi,\phi]&=\int\!ds\;\biggl[\sum_{i=1}^{2}\frac{m}{2}\,\dot{\chi}_{i}^{2}(s)-\frac{m\omega_{b}^{2}}{2}\,\chi_{i}^{2}(s)\biggr]-\int\!ds\;m\sigma\,\chi_{1}(s)\chi_{2}(s) \notag \\ &\qquad\qquad\qquad\qquad\qquad\qquad+\int\!d^{4}x\;j(x)\phi(x)+\int\!d^{4}x\;\frac{1}{2}\,\partial_{\mu}\phi\partial^{\mu}\phi\,,
\end{align}
where the current $j(x)$ takes the form $j(x)=e\sum_{i=1}^{2}\chi_{i}(t)\,\delta^{(3)}[\mathbf{x}-\mathbf{z}_{i}(t)]$. The spacetime coordinate $x$ is understood as a shorthand notation of $(t,\mathbf{x})$. The parameter $\sigma$ in the action is the coupling strength between the two idfs, while  $e$ is the coupling constant between each idf and the bath. We have written down the action to allow for the detectors to move along an arbitrary yet prescribed trajectory $\mathbf{z}_{i}(t)$. In this work we assume they stay at rest throughout.

When the initial state of the idf has a Gaussian form, the reduced density matrix of the idf can be found exactly with the help of the influence functional formalism in the closed-time path integral framework.  This enables us to obtain the full-time dynamics of the reduced system under the influence of the environment for arbitrary coupling strengths, as was done in full detail in~\cite{QEnt0}.  Here, to highlight the physics behind thermal entanglement, we opt for a simpler, more physically transparent yet no less general way, by means of the Langevin equation approach, which has been shown to be totally compatible with the reduced-density-matrix description for linear systems. For the current configuration, the Langevin equations of, say, $\chi_{1}$ is given by
\begin{align}
	 &m\,\ddot{\chi}_{1}(t)+m\omega_{b}^{2}\,\chi_{1}(t)+m\sigma\,\chi_{2}(t)\notag\\
	 &\qquad-e^{2}\int_{0}^{t}\!ds'\;\Bigl[G_{R}(\mathbf{z}_{1},s;\mathbf{z}_{1},s')\chi_{1}(s')+G_{R}(\mathbf{z}_{1},s;\mathbf{z}_{2},s')\chi_{2}(s')\Bigr]=\xi_{1}(t)\,.\label{E:eqijnk2}
\end{align}
In Eq.~\eqref{E:eqijnk2}, in addition to the restoring force  $-m\omega^{2}_{b}\chi_{1}$ and the direct coupling $m\sigma \chi_{2}(t)$ with the other idf, the {essential (most interesting)} physics is contained in the nonlocal interactions generated by the system's interaction with its environment,  and  the stochastic driving force $\xi_{1}$ which recounts both the quantum and thermal noises originating from the heat bath at the location of Detector 1.  It obeys the Gaussian statistics with $\langle\xi_{1}(t)\rangle=0$ and $\langle\xi_{1}(t)\xi_{1}(t')\rangle=e^{2}\,G_{H}(\mathbf{z}_{1},t;\mathbf{z}_{1},t')$, where  $G_{H}(x,x')=\frac{1}{2}\langle\{\phi(x),\phi(x')\}\rangle$, with $\{\,, \}$ denoting symmetrization, is the Hadamard function of the scalar field. In addition, nonzero correlation of the bath between the locations of detector 1 and 2 implies $\langle\xi_{1}(t)\xi_{2}(t')\rangle=e^{2}\,G_{H}(\mathbf{z}_{1},t;\mathbf{z}_{2},t')$. The $\langle\cdots\rangle$ can represent the ensemble average or the quantum expectation values, depending on the context.

The nonlocal expressions in \eqref{E:eqijnk2} containing the retarded Green function $G_{R}(x,x')=i\,\theta(t-t') [\phi(x),\phi(x')]$ of the scalar field,  with $[\,,]$ denoting anti-symmetrization, embrace the dissipative self-force and the history-dependent non-Markovian interaction between the two idfs as the consequences of coupling between the idfs and the bath. In particular, these nonlocal expressions are independent of the initial bath state. Essentially the stochastic forcing term and the nonlocal terms in \eqref{E:eqijnk2} capture  the overall influences from the environment. The temporal Fourier transforms of these two kernel functions $G_{H}$ and $G_{R}$ are connected via the fluctuation-dissipation relation,
\begin{align}
	 \overline{G}_{H}(\mathbf{R},\kappa)&=\coth\dfrac{\beta\kappa}{2}\,\operatorname{Im}\overline{G}_{R}(\mathbf{R},\kappa), &&{\rm where}& G(\mathbf{R},\tau)&=\int_{-\infty}^{\infty}\!\frac{d\kappa}{2\pi}\;\overline{G}(\mathbf{R},\kappa)\,e^{-i\kappa\tau}.
\end{align}
In certain contexts it signifies a balance between the energy transfer via  noise from,  and the dissipation back to,  the environment. Thus the stochastic equations of motion of $\chi_{1}$, $\chi_{2}$ describe a set of coupled, damped, driven oscillators undergoing non-Markovian dynamics.

\section{Dynamics}

The set of equations of motion for $\chi_{1}$, $\chi_{2}$ in fact can be decoupled into the center of mass (CoM) mode $\chi_{+}=(\chi_{1}+\chi_{2})/2$ and the relative mode $\chi_{-}=\chi_{1}-\chi_{2}$~\cite{QEnt0},
\begin{align}
	 \ddot{\chi}_{+}(t)+2\gamma\,\dot{\chi}_{+}(t)-2\gamma\,\frac{\theta(t-\ell)}{\ell}\,\chi_{+}(t-\ell)+\omega^{2}_{+}\,\chi_{+}(t)&=\frac{1}{m}\,\xi_{+}(t)\,,\label{E:iuwrakjd1}\\
	 \ddot{\chi}_{-}(t)+2\gamma\,\dot{\chi}_{-}(t)+2\gamma\,\frac{\theta(t-\ell)}{\ell}\,\chi_{-}(t-\ell)+\omega^{2}_{-}\,\chi_{-}(t)&=\frac{1}{m}\,\xi_{-}(t)\,.\label{E:iuwrakjd2}
\end{align}
Here the damping term and the retarded term are derived from the nonlocal expressions in \eqref{E:eqijnk2}, whose cutoff dependent component is absorbed with the bare frequency $\omega_{b}$ into a renormalized frequency $\omega$. The normal-mode frequency $\omega_{\pm}$ is then defined by $\omega_{\pm}^{2}=\omega^{2}\pm\sigma$, and the damping constant $\gamma$ by $\gamma=e^{2}/8\pi m$. The unit step function in \eqref{E:iuwrakjd1} and \eqref{E:iuwrakjd2} clearly indicates that once the idfs come into interaction with the bath at $t=0$, it takes some finite time $\ell$ for the disturbance in the field environment induced by one of the detector to reach the other detector, where $\ell$ is the separation between the two detectors (with $c=\hbar=1$). Subsequently the modified evolution of the second detector will prompt and send new bath disturbance back to the first one; this back and forth  process depends on the earlier evolutionary histories of both idfs and is thus non-Markovian.  For strong oscillator-bath coupling, measured by $2\gamma/\omega_{\pm}^{2}\ell>1$, this non-Markovian evolution can be shown to be unstable~\cite{QEnt0}. This behavior is in stark contrast to the strong coupling ($\gamma/\omega_{\pm}>1$) Markovian dynamics, such as the dynamics of the CoM mode for the two idfs in the same location \eqref{E:nckrwq1} where there is no mutual influence through the field, which tends to be overdamped. To avoid this instability, we will assume weak oscillator-bath coupling ($2\gamma/\omega_{\pm}^{2}\ell<1$) throughout. In addition, it can be shown that non-Markovianity is exponentially suppressed in the high temperature region ($\beta/\ell\ll1$) but merely algebraically in the low temperature limit ($\beta/\ell\gg1$). This implies that at high temperature (scaled by the separation), non-Markovianity is negligibly noticeable, but plays an increasingly important role in the system's nonequilibrium evolution and entanglement dynamics at low temperatures.

Since this retardation effect depends on the separation, if we place the two detectors sufficiently close to one another, we observe that the relative mode damps at a much slower rate than the CoM mode~\cite{QEnt0},
\begin{align}
	 \ddot{\chi}_{+}+4\gamma\,\dot{\chi}_{+}(t)+\widetilde{\omega}^{2}_{+}\,\chi_{+}(t)+\cdots&=\frac{1}{m}\,\xi_{+}(t)\,,\label{E:nckrwq1}\\
	 \ddot{\chi}_{-}(t)-\frac{\gamma\ell^{2}}{3}\,\dddot{\chi}_{-}(t)+\widetilde{\omega}_{-}^{2}\chi_{-}(t)+\cdots&=\frac{1}{m}\,\xi_{-}(t)\,,\label{E:nckrwq2}
\end{align}
where $\cdots$ represents the higher-order terms from the Taylor expansion of the retarded terms, and $\widetilde{\omega}_{\pm}^{2}=\omega^{2}\pm\sigma\mp\dfrac{2\gamma}{\ell}$. We see that dissipation results from the third-order time derivative for the relative mode, which is typically weaker than the counterpart for the CoM mode by the order $\widetilde{\omega}_{\pm}^{2}\ell^{2}$ in this short separation case. If this term {were} inadvertently excluded, then the relative mode {would} appear to be described by an undamped oscillator\footnote{The stochastic force $\xi_{-}$ diminishes in the short separation limit since $\xi_{-}=\xi_{1}-\xi_{2}$ and $\xi_{1}$, $\xi_{2}$ are evaluated at almost the same spacetime point.}, and the information about the initial state of this mode {would seem to} last forever, instead of being damped away. It would lead to a completely different dynamics of the relative mode. Thus these non-decaying behaviors should be more precisely understood as transients which last only within the time scale $(\gamma\widetilde{\omega}_{\pm}^{2}\ell^{2})^{-1}$.

\section{The Covariance Matrix}

The equations of motion \eqref{E:iuwrakjd1}, \eqref{E:iuwrakjd2} enable us to compute the elements of the covariance matrix, defined by
\begin{equation}
	\mathbf{V}=\frac{1}{2}\operatorname{Tr}\Bigl[\rho\bigl\{\mathbf{X},\mathbf{X}^{T}\bigr\}\Bigr]\,,
\end{equation}
where $\mathbf{X}^{T}=(\chi_{1},p_{1},\chi_{2},p_{2})$ in the case of a bipartite system and $\rho$ is the corresponding density matrix. For example, the $\mathbf{V}_{11}$ element is $\langle\chi_{1}^{2}(t)\rangle=\langle\chi_{+}^{2}(t)\rangle+\frac{1}{4}\,\langle\chi_{-}^{2}(t)\rangle$, where
\begin{align}
	 \langle\chi_{\pm}^{2}(t)\rangle&=d_{1}^{(\pm)\,2}(t)\langle\chi_{\pm}^{2}(0)\rangle+\frac{1}{m^{2}}\,d_{2}^{(\pm)\,2}(t)\langle p_{\pm}^{2}(0)\rangle\notag\\
	 &\qquad\qquad+\frac{1}{m^{2}}\int_{0}^{t}\!ds\int_{0}^{t}\!ds'\;d_{2}^{(\pm)\,2}(s)d_{2}^{(\pm)\,2}(s')\langle\xi_{\pm}(s)\xi_{\pm}(s')\rangle\,.
\end{align}
Here,  $d_{1,2}^{(\pm)}(t)$ are a special set of homogeneous solutions to \eqref{E:iuwrakjd1} and \eqref{E:iuwrakjd2}, satisfying $d_{1}^{(\pm)}(0)=1$, $\dot{d}_{1}^{(\pm)}(0)=0$, $d_{1}^{(\pm)}(0)=0$, $\dot{d}_{2}^{(\pm)}(0)=1$, and they are all equal to zero for $t<0$. The noise correlation functions can be expressed in terms of the Hadamard functions of the environment, $\langle\xi_{+}(s)\xi_{+}(s')\rangle=\dfrac{e^{2}}{2}\Bigl[G_{H}(\mathbf{0},s-s')+G_{H}(\mathbf{z}_{1}-\mathbf{z}_{2},s-s')\Bigr]$, $\langle\xi_{-}(s)\xi_{-}(s')\rangle=2e^{2}\Bigl[G_{H}(\mathbf{0},s-s')-G_{H}(\mathbf{z}_{1}-\mathbf{z}_{2},s-s')\Bigr]$ and $\langle\xi_{+}(s)\xi_{-}(s')\rangle=0$. Since the functions $d_{1,2}^{(\pm)}(t)$ in this case damp exponentially with time $t$, after the whole reduced system is fully relaxed, the terms that depend on the initial conditions $\langle\chi_{\pm}^{2}(0)\rangle$, $\langle p_{\pm}^{2}(0)\rangle$ will be negligible at late times. The other elements of the covariance matrix can be constructed likewise.

\section{Entanglement Measure}
\begin{figure}
	\centering
    \scalebox{0.45}{\includegraphics{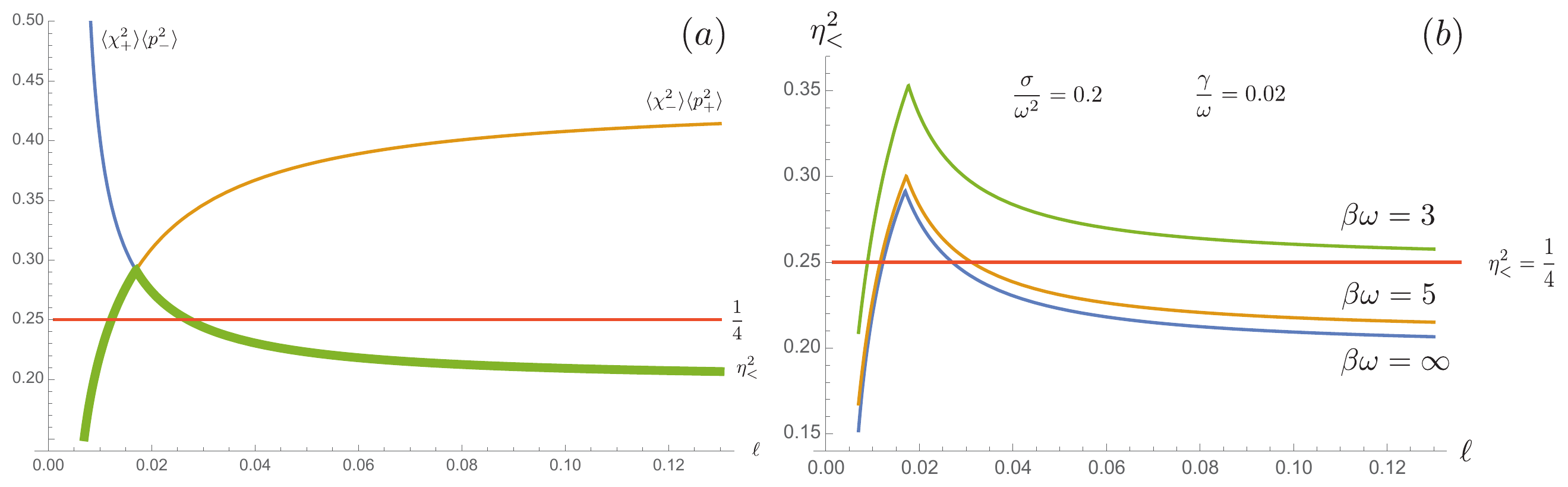}}
    \caption{(a) Typical behavior of $\eta_{<}^{2}$ as a function of separation $\ell$ between the two oscillators. An entangled state exists when $\eta_{<}^{2}<1/4$; otherwise the state is separable. There may be two different critical separations where the entanglement disappears. (b) Higher bath temperature will raise the curves upwards to make late-time entanglement harder to survive.}\label{Fi:etasml}
\end{figure}
The covariance matrix for a Gaussian continuous variable system is finite dimensional. This make it possible to construct the entanglement measures, such as negativity $\mathcal{N}(\rho)$ and logarithmic negativity $E_{\mathcal{N}}(\rho)$~\cite{Vidal,Plenio}, based on the smaller symplectic eigenvalues $\eta_{<}$ of the partially-transposed covariance matrix $\mathbf{V}^{pt}$. Negativities are calculable, and more importantly, they offer unambiguous quantification of entanglement for a symmetric two-mode Gaussian state~\cite{Eisert98,Virmani,Adesso05}. They are defined by
\begin{align}\label{E:dfkefjdkw}
	\mathcal{N}(\rho)&=\max\bigl\{0,\frac{1-2\eta_{<}}{2\eta_{<}}\bigr\}\,,&E_{\mathcal{N}}(\rho)&=\max\bigl\{0,-\ln 2\eta_{<}\bigr\}\,.
\end{align}
Entanglement occurs when $\eta_{<}<1/2$, and the degree of entanglement can be described by the negativity. In the current case, the symplectic eigenvalues $\eta_{\gtrless}$ of $\mathbf{V}^{pt}$ take on particularly neat forms
\begin{align}\label{E:qejdla}
	\eta_{<}^{2}&=\min\bigl\{\langle\chi_{+}^{2}\rangle\langle p_{-}^{2}\rangle,\,\langle\chi_{-}^{2}\rangle\langle p_{+}^{2}\rangle\bigr\}\,,&\eta_{>}^{2}&=\max\bigl\{\langle\chi_{+}^{2}\rangle\langle p_{-}^{2}\rangle,\,\langle\chi_{-}^{2}\rangle\langle p_{+}^{2}\rangle\bigr\}\,.
\end{align}
This makes interpretation of entanglement accessible.  The idea is that for the entanglement to exist, we would like to have the uncertainties of the corresponding canonical variables as small as possible in order for $\eta_{<}^{2}<1/4$. The functional form of $\eta_{<}^{2}$ will depend on the choices of the parameters $\sigma$, $\gamma$ and $\omega_{\pm}$ and its typical behavior is shown in Fig.~\ref{Fi:etasml}-(a). The structure of the late-time entanglement is much more complicated due to the interplay between different couplings and finite oscillator separation.

\section{Effective Description}
\begin{figure}
	\centering
    \scalebox{0.45}{\includegraphics{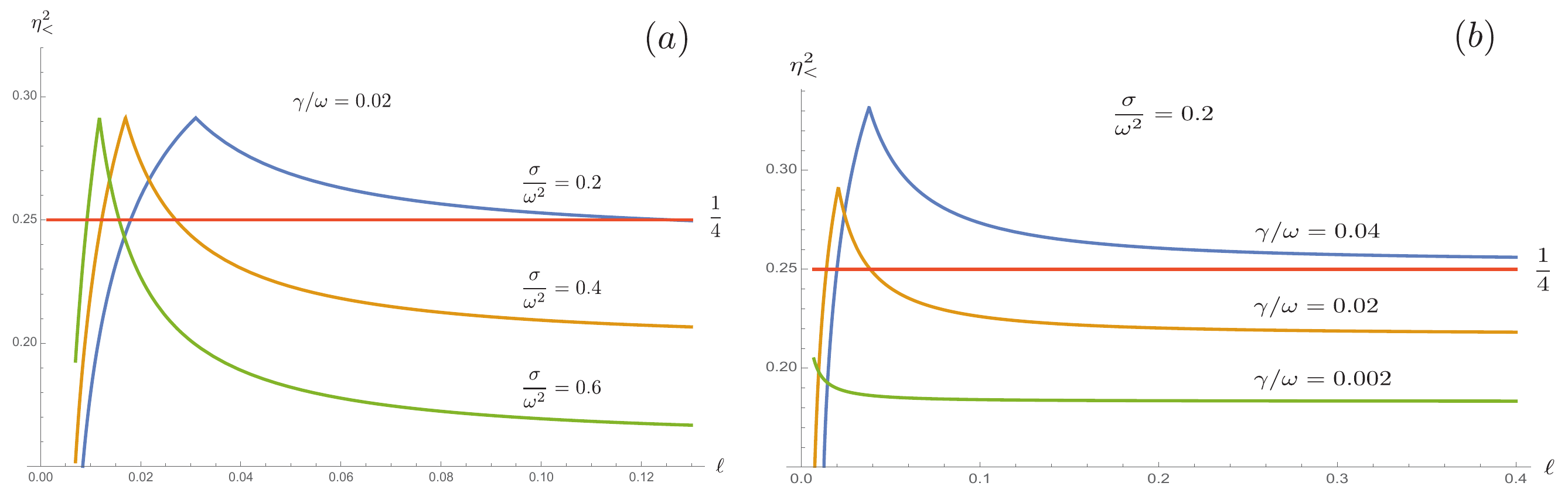}}
    \caption{The curves of $\eta_{<}^{2}$, related to negativity by \eqref{E:dfkefjdkw}, are plotted with respect to the oscillator separation for different (a) direct coupling strengths $\sigma$ and (b) oscillator-bath coupling $\gamma$. Larger $\sigma$ and smaller $\gamma$ favor the existence of late-time entanglement. These are examples at $T=0$. Finite temperature will raise all of these curves upwards, as shown in (b) of Fig.~\ref{Fi:etasml}.}\label{Fi:etaall}
\end{figure}
From the hindsights of the detailed calculations, Eqs.~\eqref{E:iuwrakjd1}, \eqref{E:iuwrakjd2} imply that at late time the normal modes can be effectively described by two uncoupled, damped, driven oscillators with the \textit{effective oscillating frequencies} $W_{\pm}^{2}=\omega^{2}\pm\sigma\mp\dfrac{2\gamma}{\ell}$ and the \textit{effective damping constants} $\Gamma_{\pm}=\gamma\pm\dfrac{\gamma}{\omega_{\pm}\ell}\,\sin\omega_{\pm}\ell$, respectively. The non-Markovian effect is still encapsulated by the expression $\gamma/\ell$. This allows us to heuristically interpret the behaviors of entanglement based on the observations that when the temperature is not too high, the uncertainty of the normal-mode canonical variables is about the order (apart from the mass scale) $\langle\chi_{\pm}^{2}\rangle\sim\mathcal{O}(W_{\pm}^{-1})$, $\langle p_{\pm}^{2}\rangle\sim\mathcal{O}(W_{\pm})$, and that at high temperatures their values are dominated by the temperature, leading to $\langle\chi_{\pm}^{2}\rangle\sim\mathcal{O}(\beta^{-1} W^{-2}_{\pm})$, $\langle p_{\pm}^{2}\rangle\sim\mathcal{O}(\beta^{-1})$. At zero temperature\footnote{The low-temperature case applies as well because the modification in uncertainties is usually algebraically small and of higher orders in $\beta^{-1}$.}, the behaviors of $\eta_{<}^{2}$ are determined by either $\mathcal{O}(W_{-}/W_{+})$ or $\mathcal{O}(W_{+}/W_{-})$, depending on $W_{+}\gtrless W_{-}$. This suggests that the curves $\langle\chi_{+}^{2}\rangle\langle p_{-}^{2}\rangle$, and $\langle\chi_{-}^{2}\rangle\langle p_{+}^{2}\rangle$ in Fig.~\ref{Fi:etasml}-(a) intersect in the vicinity of $\ell\sim2\gamma/\sigma$, which signifies a division between the relative strengths of the direct inter-oscillator coupling and the field-induced non-Markovian influence. When $\sigma>2\gamma/\ell$, which means that the direct coupling dominates over the indirect field induced effect, we have $\eta_{<}\sim\mathcal{O}(W_{-}/W_{+})$. We expect that entanglement is possibly improved by 1) stronger inter-oscillator coupling, 2) weaker oscillator-bath interaction and 3) larger separation since those condition may decrease the values of $\eta_{<}^{2}$ in the regime $\sigma>2\gamma/\ell$. On the other hand, if $\sigma<2\gamma/\ell$, the role of direct coupling is insignificant, and the dynamics is largely governed by the non-Markovian field-induced process. We have $\eta_{<}\sim\mathcal{O}(W_{+}/W_{-})$. The same arguments suggest that entanglement will be enhanced by 1) weaker inter-oscillator coupling, 2) stronger oscillator-bath interaction and 3) shorter separation. Results involving non-Markovian processes are less intuitive; however, from the previous effective description at least we easily see how they affect the uncertainties of the normal-mode variables, which determine the entanglement behaviors. These qualitative descriptions based on detailed calculations also point out two interesting facts that direct coupling and the field-induced interactions between the {idfs of the two} detectors play a competing role, and that entanglement between the subsystems can possibly be maintained at distances much longer than previously expected, when there exists direct coupling between them.

\section{Critical Temperature}

At a higher temperature, the field retardation effects vanish very rapidly as the ratio $\beta/\ell<1$, so the  dynamics of the coupled-oscillators open system becomes simpler. In the current configuration entanglement cannot survive at very high temperatures because both $\langle\chi_{\pm}^{2}(\infty)\rangle$ and $\langle p_{\pm}^{2}(\infty)\rangle$ are proportional to $\beta^{-1}$ in the high temperature limit. Their products can be easily much greater than the critical value $1/4$ of $\eta_{<}^{2}$. Thus the critical temperature for given coupling strengths and separation should fall in the intermediate range of the bath temperature. This implies neither a low- nor a high-temperature approximation can give an accurate prediction of the critical temperature but the high-temperature approximation can still offer a very reasonable upper bound for the critical temperature and thus is sufficient for our discussion on the critical temperature and its dependence on the coupling strengths.

Here we only discuss the case $\sigma>\dfrac{2\gamma}{\ell}$ because stronger inter-oscillator coupling is a necessity to counter the thermal fluctuations/excitations from the bath, so as to possibly maintain the entanglement between the oscillators at higher bath temperatures. The other range, $\sigma<\dfrac{2\gamma}{\ell}$, is less interesting since from the viewpoint of the effective frequency, entanglement can be sustained only if we require $W_{-}$ to be as large as possible to counter a large $\beta^{-1}$. It implies that $\dfrac{2\gamma}{\omega_{-}^{2}\ell}$ has to be close to 1. In such a limit the reduced system tends to be unstable. {Moreover, non-Markovianity is severely impeded at high temperature such that it becomes ineffective.} In the former case, the symplectic eigenvalue $\eta_{<}$ takes the form
\begin{equation}\label{E:zkwijsk}
	 \eta_{<}^{2}\simeq\biggl(\frac{1}{\beta\omega_{+}^{2}}+\frac{2\gamma}{\beta\ell\omega_{+}^{4}}+\frac{\beta}{12}+\cdots\biggr)\biggl(\frac{1}{\beta}+\frac{2\gamma}{\pi}\ln\frac{\Lambda}{\Lambda_{\beta}}+\cdots\biggr)\,,
\end{equation}
with $\Lambda_{\beta}$ being some number much greater than $\beta^{-1}$. Apparently $\eta_{<}^{2}$ in the high temperature limit increases quadratically with $\beta^{-1}$, meaning that thermal fluctuations dominate and introduce very large uncertainties in the canonical variables of the oscillators. This is particularly transparent in the large separation limit that $\eta_{<}^{2}$ is roughly given by $\eta_{<}^{2}\sim\dfrac{1}{\beta^{2}\omega_{+}}+\cdots$. Its value in the high temperature limit can be brought down only if $\omega_{+}$ is sufficiently large. For fixed $\omega$ this can be achieved by increasing the inter-oscillator coupling strength, but only to a certain extent. The mutual influence due to separation plays a minor role. Larger separation only minimally alleviates the detrimental effect on entanglement due to thermal fluctuations.

Suppose we extrapolate \eqref{E:zkwijsk} to the intermediate range of bath temperature, and use it to identify the critical temperature by $\eta_{<}^{2}=1/4$. We obtain
\begin{align}\label{E:dljersja}
	 \beta_{c}\sim\frac{\sqrt{6}}{\omega_{+}}+\gamma\biggl(\frac{\sqrt{6}}{\omega_{+}^{3}\ell}+\frac{9}{\pi\omega_{+}^{2}\ln\dfrac{\Lambda}{\Lambda_{\beta}}}\biggr)+\cdots\,.
\end{align}
Eq.~\eqref{E:dljersja}  gives a lower bound of $\beta_{c}$, thus equivalent to the upper bound of the critical temperature $\beta_{c}^{-1}$. This and the earlier qualitative analysis all consistently give a relation that the critical temperature should be at most about the order of the magnitude $\beta_{c}^{-1}=\mathcal{O}(\omega_{+})$, that is,
\begin{equation}
	\beta_{c}\omega_{+}=\mathcal{O}(1)\,. \label{E:Tc}
\end{equation}
Eq.~\eqref{E:dljersja} also shows that  direct coupling plays a more important role in determining the critical temperature than the other factors such as the oscillator-bath interaction strength which shows up in  $\gamma$, the oscillator separation $\ell$ and the field cutoff parameters.

\section{Conclusion}

With this, we conclude that for systems well represented by two coupled harmonic oscillator detectors interacting weakly with a common heat bath it is highly unlikely that quantum entanglement can survive at high temperatures [Eq.~\eqref{E:Tc}]. To drive up the critical temperature one should  [from Eq.~\eqref{E:dljersja}] increase the  direct coupling strength between the subsystems over larger separations.  Both statements comply with our intuitions.  This settles the question we raised in the beginning for nonequilibrium (Case B) systems with time-independent coupling in a common thermal environment. \\ 


\section*{References}

\begin{thebibliography}{99}

\bibitem{GalvePRL}  F. Galve, L.A. Pach\'on and D. Zueco, ``Bringing entanglement to the high temperature limit'',  Phys. Rev. Lett. \textbf{105}, 180501 (2010).

\bibitem{EP} A. F. Estrada and L. A. Pach\'on, ``Quantum limit for driven linear non-Markovian open-quantum-systems", [arXiv:1411.3382].

\bibitem{Vedral} V. Vedral, ``Quantum physics: Hot entanglement'', Nature \textbf{468}, 769 (2010).

\bibitem{AndWin} J. Anders, and A. Winter, ``Entanglement and separability of quantum harmonic oscillator systems at finite temperature'', Quantum Inf. Comput. \textbf{8}, 0245 (2008).

\bibitem{Anders} J. Anders, ``Thermal state entanglement in harmonic lattices'', Phys. Rev. A \textbf{77}, 062102 (2008).

\bibitem{HotEnt2}  J.-T. Hsiang and B. L. Hu, `` Quantum entanglement at high hemperatures? II. Bosonic systems in nonequilibrium steady state",  [arXiv:1503.03587].

\bibitem{HHNESS}  J.-T. Hsiang and B. L. Hu, ``Nonequilibrium steady state in open quantum systems:  influence action, stochastic equation and power balance'', [arXiv:1405.7642].

\bibitem{qos}  H.-P. Breuer and F. Petruccione, {\sl The Theory of Open Quantum Systems} (Oxford University Press, Oxford, 2003).

\bibitem{CH08} E. Calzetta and B. L. Hu, {\sl Nonequilibrium Quantum Field Theory} (Cambridge University Press, Cambridge, 2008).

\bibitem{QEnt0} J.-T. Hsiang and B. L. Hu,  ``Distance and coupling dependence of entanglement in the presence of a quantum field",  [arXiv:1505.03007].

\bibitem{LH09} S. Y. Lin and B. L. Hu, ``Temporal and spatial dependence of quantum entanglement from a field theory perspective'', Phys. Rev. D \textbf{79}, 085020 (2009).

\bibitem{Vidal} G. Vidal, and R. F. Werner, ``A computable measure of entanglement'', Phys. Rev. A \textbf{65}, 032314 (2002).

\bibitem{Plenio} M. B. Plenio, ``The logarithmic negativity: A full entanglement monotone that is not convex'', Phys. Rev. Lett. \textbf{65}, 95 (2005).

\bibitem{Eisert98} J. Eisert and M. B. Plenio, ``A comparison of entanglement measures'', J. of Mod. Opt. \textbf{46}, 6 (1998).

\bibitem{Virmani} S. Virmani and M. B. Plenio, ``Ordering states with entanglement measures'', Phys. Lett. A {\bf 268}, 31 (2000).

\bibitem{Adesso05} G. Adesso and F. Illuminati, ``Gaussian measures of entanglement versus negativities: ordering of two-mode Gaussian states'', Phys. Rev. A \textbf{72}, 032334 (2005).

\end{thebibliography}
\end{document}